\documentclass[aps,twocolumn,prl,superscriptaddress,amsmath,amssymb,amsfonts,nofootinbib]{revtex4-1} 
 \usepackage{graphicx}
 \usepackage{calc}
 \usepackage{xspace}
 \usepackage[T1]{fontenc}
 \usepackage{setspace}
 \usepackage{calligra}
 \usepackage{color}
 \usepackage{soul}
\usepackage[framemethod=TikZ]{mdframed}
\mdfsetup{frametitlealignment=\center}
\mdfdefinestyle{MyFrame}{%
    linecolor=red,
    outerlinewidth=2.5pt,
    roundcorner=20pt,
    innertopmargin=\baselineskip,
    innerbottommargin=\baselineskip,
    innerrightmargin=20pt,
    innerleftmargin=20pt,
    backgroundcolor=gray!50!white}

\usepackage[left=0.5in,right=.5in,top=1in,bottom=1in]{geometry}
\DeclareMathAlphabet{\mathcalligra}{T1}{calligra}{m}{n}
\DeclareFontShape{T1}{calligra}{m}{n}{<->s*[2.2]callig15}{}

\usepackage[caption=false]{subfig}
\usepackage{ORI_Group_style}

\newcommand{\hsig}{\hat{\sigma}}
\newcommand{\D}{\Delta}
\newcommand{\G}{\Gamma}

\newcommand{\g}{\gamma}

\newcommand{\bG}{\overline{\Gamma}}
\newcommand{\hrho}{\hat{\rho}}
\newcommand{\hmu}{\hat{\mu}}
\newcommand{\Ham}{\hat{H}_A}
\newcommand{\hS}{\hat{S}}

\newcommand{\mL}{\mathcal{L}}
\newcommand{\bg}{\overline{g}}

\newcommand{\XX}{\mathbf{X}}
\newcommand{\YY}{\mathbf{Y}}

\newcommand{\mP}{\mathcal{P}}
\newcommand{\mQ}{\mathcal{Q}}

\begin{document}
\title{Cooperative Effects in Closely Packed Quantum Emitters with Collective Dephasing}

\author{B. Prasanna Venkatesh}
\affiliation{Institute for Quantum Optics and Quantum Information of the
Austrian Academy of Sciences, A-6020 Innsbruck, Austria.}
\affiliation{Institute for Theoretical Physics, University of Innsbruck, A-6020 Innsbruck, Austria.}
\author{M. L. Juan}
\affiliation{Institute for Quantum Optics and Quantum Information of the
Austrian Academy of Sciences, A-6020 Innsbruck, Austria.}
\affiliation{Institute for Experimental Physics, University of Innsbruck, A-6020 Innsbruck, Austria.}
\author{O. Romero-Isart}
\affiliation{Institute for Quantum Optics and Quantum Information of the
Austrian Academy of Sciences, A-6020 Innsbruck, Austria.}
\affiliation{Institute for Theoretical Physics, University of Innsbruck, A-6020 Innsbruck, Austria.}
\allowdisplaybreaks[2]
\begin{abstract}
In a closely packed ensemble of quantum emitters, cooperative effects are typically suppressed due to the dephasing induced by the dipole-dipole interactions. Here, we show that by adding sufficiently strong collective dephasing cooperative effects can be restored. In particular, we show that the dipole force on a closely packed ensemble of strongly driven two-level quantum emitters, which collectively dephase, is enhanced in comparison to the dipole force on an independent non-interacting ensemble. Our results are relevant to solid state systems with embedded quantum emitters such as colour centers in diamond and superconducting qubits in microwave cavities and waveguides.
\end{abstract}
\maketitle

A collection of two-level quantum emitters (TLEs) with sub-wavelength average separations can show remarkable cooperative behaviour like superradiant emission~\cite{Agarwal74,Gross82,Ficek87}. The study of optical response in such systems has predominantly been restricted to the emission properties or the propagation of light within the TLE ensemble. This is because the systems usually considered in the early days~\cite{Gross82,Malcuit87}, as well as in some recent works~\cite{Bachelard11,Guerin16,AraujoKemp16,Jennewein16}, are gaseous clouds of atoms. With the advent of artificial atoms in solid state systems, \eg~quantum dots~\cite{Brandes05}, superconducting qubits~\cite{Nissen13,Mlynek14} and colour centers in diamond~\cite{NVreview,Juan16,Bradac16}, it is now possible to study the impact of cooperative effects on other aspects of the optical response. In particular, a recent experiment~\cite{Juan16} studied the dipole force on optically trapped nanodiamonds containing a high density of Nitrogen vacancy (NVs) centers. An intriguing result in~\cite{Juan16} was that the observed dipole force originating from the emitters could not be correctly accounted for by considering the emitters to respond independently. 

In this work, we focus on cooperative effects in a small and closely packed ensemble of TLEs subject to strong coherent driving and collective dephasing. In particular, we show that the dipole force on such an ensemble can be larger than on an equivalent one where each TLE spontaneously emits \emph{independently}. For the emitter separations that we consider here, the dipole-dipole interaction can be larger than the line-width of the individual emitters. Furthermore, spontaneous emission is not perfectly collective. In this situation, one typically expects cooperative effects to be suppressed \cite{Gross82,Ficek83}. Here, we show that the combination of strong driving and large collective dephasing can restore cooperative effects, even in the presence of dipole interaction shifts and non-collective spontaneous emission. While there have been previous studies of cooperative effects with strong driving fields~\cite{FicekTanasReview02,AminCordes78,Agarwal79,Agarwal80,Drummond80,Carmichael82,Campagno82}, the role of collective dephasing has received less attention~\cite{Nissen13,Juan16}. In the context of Quantum Information, collective decoherence in general, and collective dephasing in particular, has been studied both theoretically \cite{Palma96,Duan98,Lidar03} and experimentally \cite{Kwiat00,Kielpinski01}. There, particular attention was paid to the existence and robustness of so called decoherence free sub-spaces (DFS) under collective dephasing~\cite{Lidar03}. Moreover, recent studies \cite{Lanyon13,Carnio15} have shown that collective dephasing could also be used as a resource to generate strong but separable correlations. {We note that, due to their promise as a general passive strategy to protect quantum resources from noise, the study of DFS continues to be an active area of research, see \cite{Lidar12} for a recent review on the theoretical aspects and \cite{DFSexpts} for experimental implementations. Of late, novel applications of DFS, such as generation of arbitrary photonic states \cite{Tudela15} and universal quantum computation in waveguide QED \cite{Tudela16} as well as quantum repeaters with trapped ions, \cite{Zwerger17} have also been proposed.}

Let us consider a collection of $N$ identical TLEs with resonance frequency $\w_0 \equiv ck_0 \equiv 2\pi c/\lambda_0$. The matrix element of the dipole moment operator $\hat{\dd}$ is given by $\sandwich{e}{\hat{\dd}}{g} \equiv \epsb_a d$, where $\ket{g}$ ($\ket{e}$) denotes the ground (excited) state and $\abs{\epsb_a}=1$. For simplicity, we assume $\epsb_a$ and $d$ to be real. The TLEs, each with position $\rr_m$ ($m=1,\ldots,N$), are fixed on a background matrix with center-of-mass position $\xx$, namely $\rr_m' \equiv \rr_m - \xx$ for all $m$ is a constant of motion. We consider the TLEs to be driven by a classical electromagnetic field of the form  $\EE(\rr,t) = \EE(\rr) \cos\pbra{\omega_d t} \equiv \epsb_d E_0 f(\rr) \cos\pbra{\omega_d t}$, where $\epsb_d$ and $E_0$ are real and $\abs{\epsb_d}=1$. We assume that $f(\rr_m) \approx \ff(\xx)$ for all $m$. In a frame rotating with the drive and assuming the rotating wave approximation, the hamiltonian describing the interaction of the identical TLEs with $\EE(\rr,t)$ is  given  by
\begin{align}
\Ham \equiv \frac{\hbar\W(\xx)}{2} \hS^x - \frac{\hbar\Delta}{2} \hS^z \label{eq:Hamdriv}.
\end{align}
Here, $\W(\xx) \equiv-2 d E_0 (\epsb_a \cdot \epsb_d) f(\xx)/\hbar$ is the Rabi frequency and $\Delta \equiv \w_d-\w_0$ the detuning. Hereafter, we use the following notation for the spin operators:  $\hsig^{\alpha}_m$ denotes the Pauli matrix (for $\alpha = {x,y,z}$) and the ladder operator (for $\alpha = \pm$) of the $m$th TLE, collective operators are denoted by $\hS^{\alpha} \equiv \sum_{m=1}^N \hsig^{\alpha}_m$. Apart from the dynamics induced by the interaction with the external driving, we assume the TLEs to experience collective dephasing as well as dipole-dipole interaction and spontaneous emission due to the interaction with free electromagnetic field modes in the vacuum state. The overall dynamics of such an ensemble of identical TLEs can  be then described by the master equation $\dot \hrho =  \mathcal{L}\hat \rho \equiv \pare{ \mathcal{L}_H + \mathcal{L}_\Gamma + \mathcal{L}_\gamma} \hat \rho$, where~\cite{Agarwal74,Gross82,Ficek87}
\begin{align}
\mathcal{L}_H \hat \rho & \equiv \frac{1}{\im \hbar}[\Ham +\Hop_I,\hrho ], \\
\mathcal{L}_\Gamma \hat \rho & \equiv \sum_{mn} \Gamma_{mn} \pare{ 2\hsig_m^- \hrho \hsig_n^+ - \hsig_m^+\hsig_n^-\hrho - \hrho \hsig_m^+\hsig_n^-  }, \label{eq:collemit}\\
\mathcal{L}_\gamma \hat \rho & \equiv - \frac{\gamma_c}{4} [\hS^z,[\hS^z,\hrho ]].
\end{align}
The term $\mathcal{L}_H$ describes coherent dynamics given by the interaction with the external field, \eqnref{eq:Hamdriv}, and the dipole-dipole interaction given by $\Hop_I \equiv \sum_{m\neq n} \hbar g_{mn}  \hsig_m^+\hsig_n^- $, where \cite{Kraemer16}
\be \label{eq:wdd}
\begin{split}
g_{mn} \equiv  -\frac{3 \G}{4} &\left [\pbra{1-\cos^2 \theta_{mn}}\frac{\cos{\xi}}{\xi} \right . \\
 & \left .   - \pbra{1-3\cos^2 \theta_{mn}} \left(\frac{\cos{\xi}}{\xi^3}+ \frac{\sin{\xi}}{\xi^2}\right) \right] .
\end{split}
\ee 
Here, $\xi=k_0 r_{mn} \equiv k_0 \abs{\rr_m - \rr_n}$, and $ \cos \theta_{mn} \equiv (\rr_m - \rr_n) \cdot \epsb_a /r_{mn}$. The term $\mathcal{L}_\Gamma$ describes the spontaneous emission of the TLEs with correlated emission rates given by
\be \label{eq:gamdd}
\begin{split}
 \G_{mn} \equiv  \frac{3\G}{4} &\left [\pbra{1-\cos^2 \theta_{mn}} \frac{\sin(\xi)}{\xi} \right . \\
 & \left .-\pbra{1-3\cos^2 \theta_{mn}}\left(\frac{\sin{\xi}}{\xi^3}-\frac{\cos{\xi}}{\xi^2}\right) \right ]  .
\end{split}
\ee
The diagonal term $\G_{mm} \equiv \Gamma/2 = 2 d^2 \w_0^3/(3\hbar c^3)$ is the individual spontaneous emission rate of the TLEs. The term $\mathcal{L}_\gamma$ describes collective dephasing with a rate given by $\gamma_c$. {It is convenient to introduce the rate $\bar g \equiv \sum_{n \neq 1} \abs{g_{1n}}$, which parameterizes the strength of the dipole-dipole interaction, and $\bG \equiv  \sum_{n \neq 1} \G_{1n}/(N-1)$, which parameterizes the cooperativity of the spontaneous emission}. The physical origin of collective dephasing is left unspecified in the theoretical treatment here. Note that it can, for instance, arise via correlated magnetic field fluctuations for ions \cite{Lanyon13} or due to interactions with phononic baths in the case of colour centers \cite{Palma96,Fu2009,Albrecht13} (see \cite{Supplement} for additional details regarding such situations).

\begin{figure}
\centering
\includegraphics[width=8.6cm,height=10.7cm]{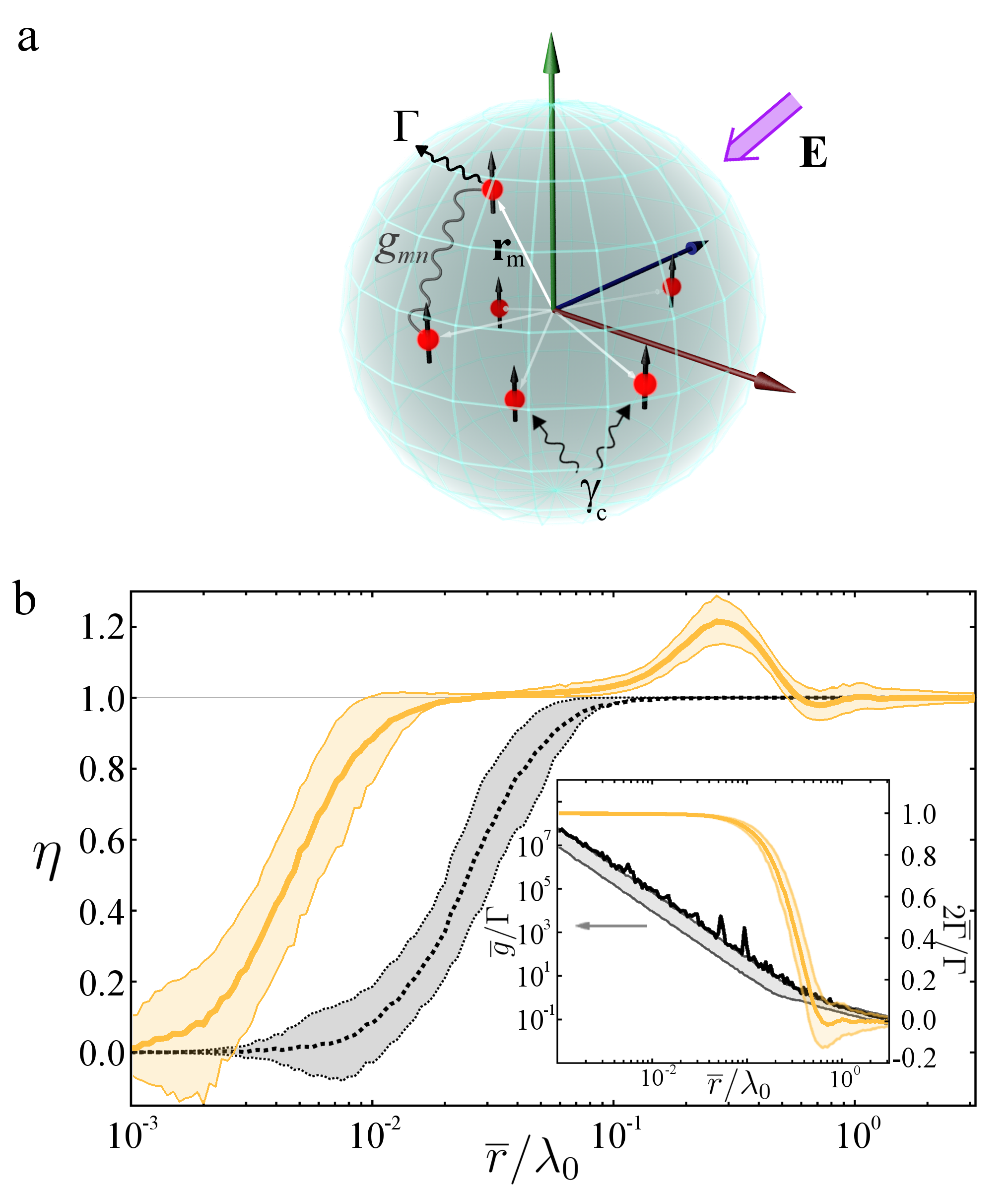}
\caption{ (a) Schematic illustration of a driven ensemble of randomly distributed identical TLEs in a 3D volume.  (b) $\eta$ is plotted as a function of $\bar r/\lambda_0$ for $N=6$. The thick lines are the mean $\eta$ over 1000 random distributions and the shaded areas represent regions where 68$\%$ of the values for $\eta$ lie (see~\cite{Supplement} for further details). Solid (dotted) line corresponds to $\g_c/\G = 1.3 \times 10^4 $ ($\gamma_c=0$) for $\W_0/\Gamma = 10^3$. Inset plots the mean and the 68$\%$ confidence interval of $\bg/\Gamma$ and $2 \bar \Gamma/\Gamma$ as a function of $\bar r/\lambda_0$.}
\label{fig:fig_N6}
\end{figure}
We are interested in the closely packed regime defined by $k_0 r_{mn} \leq 1$ for any pair of TLEs. While in this regime the spontaneous emission is predominantly collective, namely $\bar \Gamma \lesssim \Gamma/2$, the strong dipole-dipole interaction $\bar g \gg \Gamma$ typically suppresses any cooperative effect. However, we show below that strong collective dephasing $\gamma_c  \gg \Gamma$ together with strong driving $\Omega(\xx) \gg  \Gamma$ can recover cooperative effects. In particular, we concentrate on the steady state value of $\hat S^x$, namely $\avg{\hat S^x} \equiv \tr [\hat S^x \hat \rho_{s}]$, where $\mathcal{L} \hat \rho_{s} =0$. This is related to the dipole force exerted by the driving field to the matrix hosting the TLEs. Indeed, assuming that the motion of the background matrix in the applied field is slow compared to the emitter dynamics, the dipole force is given by
\begin{align}
    \FF_{\mathrm{dp}} = -\frac{\hbar \nabla \W(\xx)|_{\xx_0}}{2}  \avg{\hS^x}  \label{eq:fdpgen},
\end{align}
where $\xx_0$ is the equilibrium position of the matrix. For an ensemble of $N$ independent TLEs, namely with $g_{mn}\equiv 0$ and $\G_{mn}\equiv 0$  (for $m \neq n$), one has that
\be \label{eq:indemitpol}
\begin{split} 
 \avg{\hS^x}_\text{ind} &= N \frac{4  \Delta \Omega_0\Gamma}{\Gamma\left(4\Delta^2+\gamma_{\perp}^2\right)+2\Omega_0^2\gamma_{\perp}} \\
 & \leq N \frac{ \Omega_0 \Gamma}{\sqrt{\Gamma \gamma_{\perp}\left(\Gamma \gamma_{\perp}+2\W_0^2\right)}} .
 \end{split}
\ee 
Here, $\Omega_0 \equiv \Omega(\xx_0)$, $\gamma_{\perp} \equiv (\Gamma+2\gamma_c)$, and the upper bound is achieved at the optimal detuning 
\be \label{eq:detuning}
\Delta_0 \equiv -\sqrt{\frac{\gamma_{\perp}\left( \Gamma \gamma_{\perp} + 2 \W_0^2\right)}{4\Gamma}}.
\ee 
We are interested in the parameter $\eta \equiv \avg{\hS^x}/ \avg{\hS^x}_\text{ind}$ evaluated at $\Delta= \Delta_0$ (note that $\avg{\hS^x}$ is not maximized at $\Delta_0$). In particular, we refer to situations when $\eta > 1$ as {\em cooperative enhancement} (CE). We remark that since we are interested in closely packed ensembles, we do not consider variations of the Rabi frequency with emitter locations $\rr_m$, which can also lead to interesting modifications of collective effects \cite{Zoubi10,AsenjoGarcia17}. 

Let us consider $N$ identical TLEs, randomly positioned in a three dimensional volume with an average separation given by $\bar r \equiv \sum_{m>n} r_{mn}/N$, see \figref{fig:fig_N6}a. We generate multiple random configurations at a fixed $\bar r$ using the procedure described in \cite{Damanet16} (see \cite{Supplement} for details) and numerically calculate  $\hat \rho_s$ in each case using \cite{Qutip13}. The average $\eta$ over 1000 configurations is plotted in~\figref{fig:fig_N6}b as a function of $\bar r/\lambda_0$ for $N=6$  with (solid line) and without (dotted line) collective dephasing. The shaded regions correspond to the interval where a majority, 68$\%$, of the values for $\eta$ lie. In the absence of collective dephasing ($\g_c = 0$), there is no CE ($\eta \leq 1$). However, in the presence of strong collective dephasing ($\gamma_c/\Gamma\approx10^4$) and strong driving ($\Omega_0/\Gamma=10^3$) there is a range of separations $\bar r$ where there is CE ($\eta>1$). This is the main finding of this work. In the inset of~\figref{fig:fig_N6}b, the parameters $\bar g/\Gamma$ and $2 \bar \Gamma/\Gamma$ as a function of $\bar r$ (averaged over 1000 configurations) are plotted. Note that CE vanishes both at large average separations due to the non-collective nature of spontaneous emission and at small distances due to the large dipole-dipole interaction. 

\begin{figure}
\centering
\includegraphics[width=0.9\linewidth]{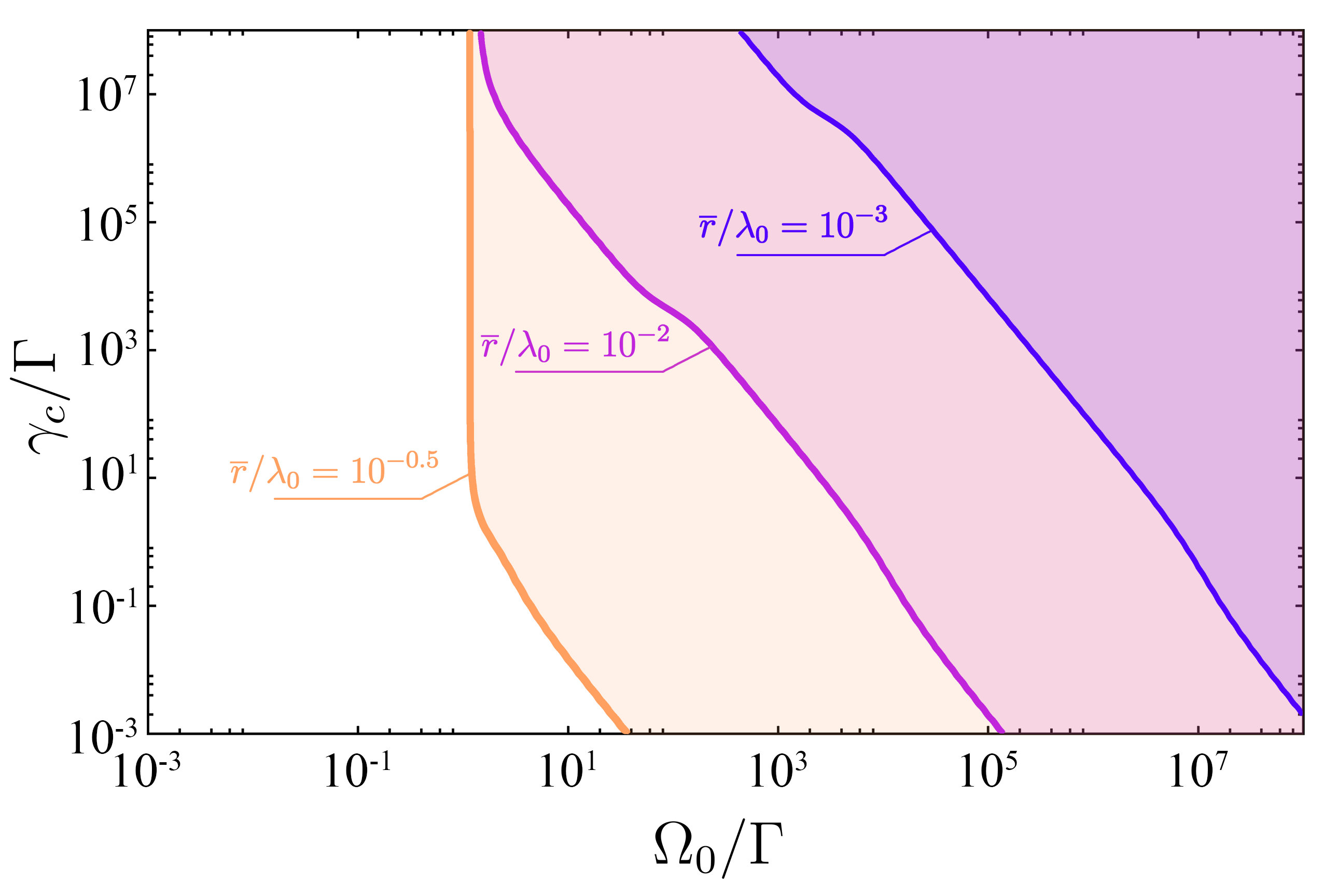}
\caption{Regions of CE for $N=2$ in the plane $(\W_0/\Gamma,\g_c/\Gamma)$ for different $\bar r/\lambda_0$. The contour line corresponds to $\eta = 1$.}
\label{fig:fig_contour}
\end{figure}

Let us analytically support these statements for the simplest $N=2$ case~\cite{Ficek83}. The hamiltonian including the dipole interaction is
\be 
\mathcal{L}_H \hat \rho  \equiv \frac{1}{2 \im }[\W_0 \hS^x - (\Delta_0+\bar g) \hS^z + 2 \bar g \hS^+ \hS^-,\hrho  ].
\ee 
In this case, the hamiltonian is collective and commutes with $\hat{\mathbf{S}}^2$. In contrast, in the finite emitter separation regime, where the so-called small sample limit~\cite{Dicke54,Agarwal74,AminCordes78,Gross82,FicekTanasReview02} can not be used, the spontaneous emission~\eqnref{eq:collemit} is still non-collective \ie $\chi \equiv 2 \bG/ \G <1$. Following~\cite{Ficek83} and as shown in~\cite{Supplement}, one can analytically calculate $\eta$. \figref{fig:fig_contour} plots the CE region ($\eta>1$) in the plane $(\W_0/\Gamma, \g_c/\Gamma)$ for different $\bar r$. Note that CE requires both large dephasing and large driving. Furthermore, from the lengthy analytical expression for $\eta$, one can obtain that in the limit of large dephasing $\eta$ reads
\begin{align}
 \lim_{\gamma_c/\Gamma \rightarrow \infty} \pare{\eta-1} &\sim \frac{\Gamma}{\gamma_c}\frac{\chi}{2  \pare{1 + \chi}}\pare{\frac{\Omega_0^2}{\Gamma^2} - 1 -\chi }.\label{eq:etagamcneq0}
\end{align}
In this limit, CE ($\eta>1$) requires sufficiently large driving $\Omega_0 > \sqrt{\G^2 + 2\bG \G}$. Alternatively, one can also show that in the limit of no dephasing ($\gamma_c=0$) and large driving, $\eta$ reads
\begin{align}
  \lim_{\Omega_0/\Gamma \rightarrow \infty, \gamma_c=0} \pare{\eta-1} &\sim - \frac{\Gamma^2}{\Omega_0^2} \frac{1}{8} \pare{ \chi^2 + 2 \chi  + \frac{4 \bar g^2}{\Gamma^2}  }\label{eq:etagamc0},
\end{align}
and hence there is no CE ($\eta<1$). This is in agreement with previous studies of resonance fluorescence \cite{AminCordes78,FicekTanasReview02,Agarwal79,Agarwal80} at strong driving, which results from the increased occupation of the bright Dicke subspace in the small sample limit. It was shown in~\cite{Ficek83} that when $\chi <1$ such an enhancement is absent. Interestingly, we claim here that large collective dephasing can restore cooperative effects.

Returning to \figref{fig:fig_N6}b, $N=6$, we observe that there is an optimal separation distance $\bar r/\lambda_0$ where $\eta$ reaches a maximum. The dipole-dipole interactions for $N>2$ do not conserve permutation symmetry in general. As a result, large dipole interactions induce strong local dephasing apart from providing energy shifts that prevent the TLEs to be polarized at the chosen detuning $\Delta_0$. We see in~\figref{fig:fig_N6}b that in regions with CE the dipole interactions satisfy $\g_c \gg \bg \sim \G$. In this manner we can understand the maximum $\eta$ in \figref{fig:fig_N6}b as occurring at separations where the dipole interactions are small enough to not dephase the collective behaviour fostered by the cooperative spontaneous emission and collective dephasing. This statement can be quantified by noting that the product $g_{12}(1-2\G_{12}/\G)$, for a pair of TLEs with parallel moments and $\theta_{12} = \pi/2$, is locally minimized at $r_{12} \approx 0.2 \lambda_0$. This is consistent with the position of the peak in~\figref{fig:fig_N6}b. Furthermore, this also agrees with the observation that the location of the peak remains in the same density region for various ensemble sizes (see \figref{fig:fig_N4_to_10}) and driving strengths $\Omega_0$ (see \cite{Supplement}).

\begin{figure}
\centering
\includegraphics[width=8.6cm,height=5.4cm]{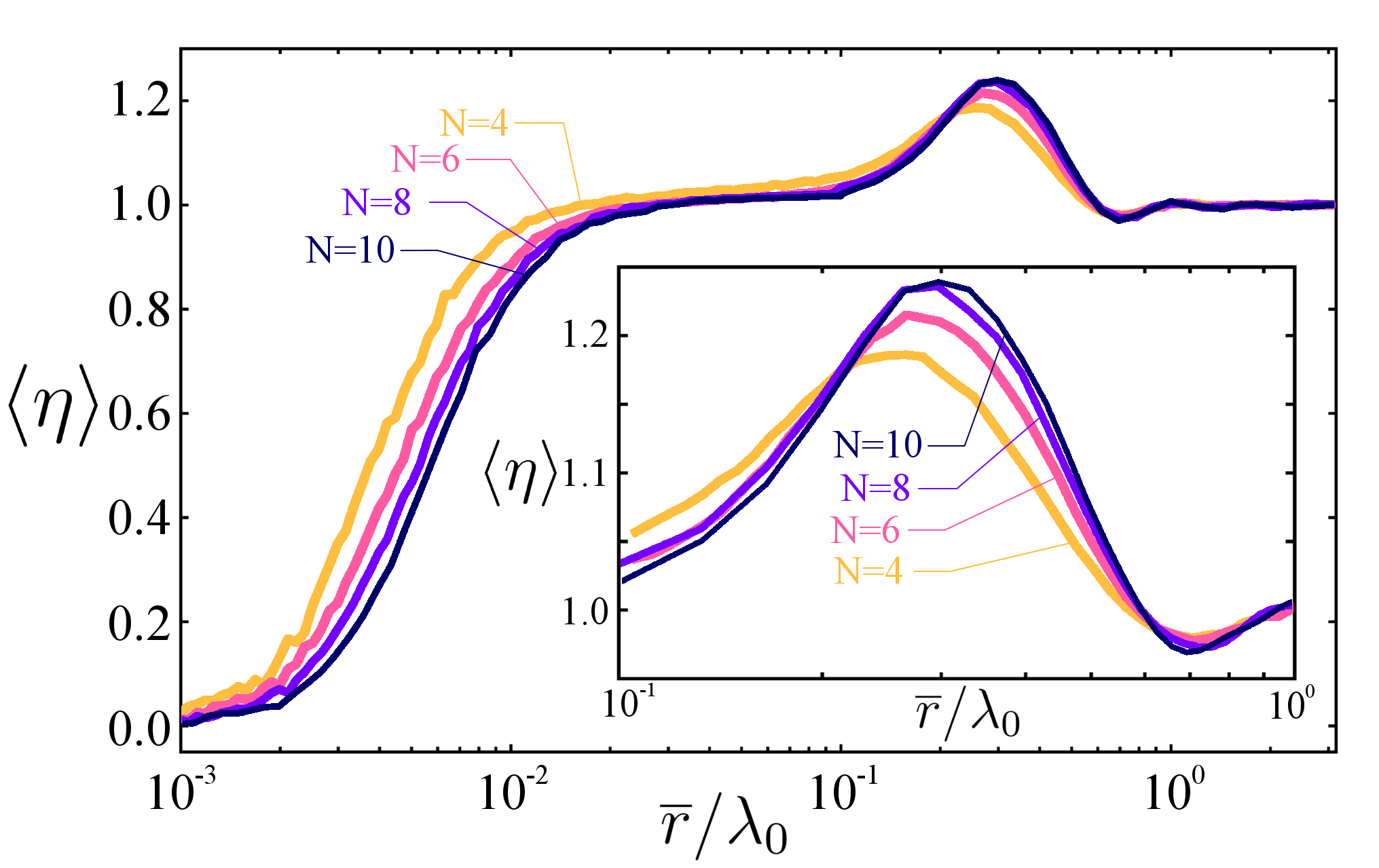}
\caption{ $\avg{\eta}$ is plotted as a function $\bar r/\lambda_0$ for different $N$ using a rate equation approximation for  $\bar r/\lambda_0 <10^{-1}$ (averaged over 400 random configurations) and using a trajectory method simulation (averaged over 200 configurations) $\bar r/\lambda_0 > 10^{-1}$. Inset is the zoomed-in region of CE.}
\label{fig:fig_N4_to_10}
\end{figure}
For $N>6$, an exact numerical calculation of $\hat \rho_s$ becomes rapidly intractable with standard resources. State-of-the art approximation methods in the field, such as Holstein-Primakoff~\cite{Jenkins12,Bettles16a,Lee16} or extended mean-field~\cite{Kraemer15}, are not valid in the strong driving ($\Omega_0 \gg \Gamma$) and strong dipole-dipole interaction ($\bar g \gg \Gamma$) regime respectively. 
The separation of time scales between the spontaneous emission and collective dephasing $\g_c \gg \G$ also makes trajectory methods ~\cite{Clemens03} unsuitable. For small emitter separations up to $\bar{r}/\lambda_0 \sim 10^{-1}$, we find that a numerical diagonalization of the hamiltonian  $\Ham+\Hop_I$  followed by a secular approximation to convert the master equation to a rate equation in the dressed basis~\cite{Dalibard85} allows us to go up to $N=10$ emitters. For $\bar{r}/\lambda_0 \gtrsim 10^{-1}$ the weak dipole-dipole interactions do not appreciably lift the degeneracy between the Dicke states with the same angular momentum projection $m$ but different total angular momentum $S$, rendering the dressed approach invalid. In this regime we proceed as follows. The Liouvillian of the master equation can be written as
$\mathcal{L} = \mathcal{L}_0  + \mathcal{L}_1 $, where $\mathcal{L}_0 \hat \rho \equiv \mathcal{L}_\gamma \hat \rho + \im \Delta_0[\hat S^z, \hat \rho]/2$ and $\mathcal{L}_1 \equiv \mathcal{L}_\Gamma \hat \rho + (\im \hbar)^{-1}[\Hop_I + \hbar \Omega_0 \hat S^x/2, \hat \rho] $. In most of the regime where $\eta>1$, one has that $\gamma_c$ and $\D_0$  are much larger than $\bar \Gamma$,  $\bar g$,  and $\Omega_0$. Indeed, note that $\D_0 \approx \Omega_0 \sqrt{\gamma_c/\Gamma}$ when $\W_0^2 \gg \G \g_c$, which according to \eqnref{eq:etagamcneq0}, is required to have an appreciable value of $\eta-1>0$. Under these assumptions, $\mathcal{L}_0$ describes faster dynamics than $\mathcal{L}_1$, and hence, one can adiabatically eliminate~\cite{Cirac92,Lesanovsky13}  the fast dynamics. This leads to  an effective master equation in the dark subspace of $\mathcal{L}_0$, namely for  states $\hat \mu$ such that $\mathcal{L}_0 \hat \mu = 0$. As shown in~\cite{Supplement}, the effective master equation is given by
\be  \label{eq:effmastermu}
\dot \hmu = \frac{1}{\im \hbar} \com{\Hop_I}{\hmu} + \mathcal{L}_{\G} \hmu -\frac{\kappa}{4} ( \coms{\hS^{-}}{\coms{\hS^+}{\hmu}}+\coms{\hS^{+}}{\coms{\hS^-}{\hmu}}).
\ee 
Here $\kappa \equiv \W_0^2 \g_c/\pbra{\g_c^2+\D_0^2}$ is of the order of $\Gamma$ in the assumed parameter regime. \eqnref{eq:effmastermu} can be conveniently solved numerically via trajectory unravellings~\cite{Clemens03}. 

In \figref{fig:fig_N4_to_10}, the results for the averaged $\eta$ over multiple random configurations, $\avg{\eta}$, for $8$ and $10$ TLEs are presented. For $\bar{r}/\lambda_0 < 10^{-1}$, $\avg{\eta}$ was calculated by averaging over $400$ random configuration with steady states calculated using the rate equation method. For $\bar{r}/\lambda_0 > 10^{-1}$, the mean was calculated over $200$ configurations from the steady state solutions determined by averaging over 500 trajectories each \cite{Qutip13}.

The statistical distribution of $\eta$, and checks to ensure that the approximate methods used for $N=8,10$ in \figref{fig:fig_N4_to_10} agree with the exact results for $N\leq 6$ are presented in \cite{Supplement}. From the inset in \figref{fig:fig_N4_to_10}, we see that while $\avg{\eta}$ increases with $N$ in the CE region up to $N=8$, there is no appreciable gain for $N=10$. In \cite{Supplement}, we also demonstrate CE for an equidistant circular arrangement of TLEs. Since dipole-dipole interactions are permutation symmetric in this case, we find a significant CE also at smaller separations than in the random arrangement. We remark that the choice of detuning $\D_0$ maximizes $\avg{S^x}$ for independent emitters and that hence, an optimized choice for the collective case can certainly lead to even larger $\eta$. A more systematic study of the CE as a function of different arrangements of TLEs~\cite{Kraemer16}, as well as developing efficient methods to numerically and analytically address larger number of TLEs is left for future work.

We conclude with some remarks relating our findings to the recent studies of dipole force on colour centres embedded in nano-diamonds~\cite{Juan16,Bradac16}. In the experiment~\cite{Juan16}, the underlying mechanism for the large dephasing at room temperature is mediated via phonon interactions~\cite{Fu2009} and consequently changes rapidly with the temperature of the lattice. Since we have demonstrated that the presence of large collective dephasing is crucial to the observation of the enhanced dipole force, this raises the prospect of repeating the experiment~\cite{Juan16}, or even life-time measurements in~\cite{Bradac16}, at lower temperatures. At lower temperatures, the dephasing will be reduced which should  lead to a strong modification or even suppression of collective effects. This is counter-intuitive to the study of collective effects in atomic systems.  In connection to the proposal for levitated optomechanics with nano-diamonds~\cite{Juan16a} (as well as other proposals~\cite{Xuereb12} concerning collective effects in optomechanics), it is interesting to explore if collective effects in dense ensembles lead to polarizabilities comparable or greater than the bulk polarizability of the embedding medium. A promising direction for further research is to explore other scenarios where large collective dephasing restores cooperative effects. Remarkably, in systems such as superconducting qubits~\cite{Nissen13,Mlynek14} where the collective dephasing can be externally controlled, this could allow to observe cooperative effects even in the presence of inhomogeneities and dipole shifts.

This work is supported by the Austrian Federal Ministry of Science, Research,and Economy (BMWFW). We acknowledge discussions with J. I. Cirac, J. J. Garcia-Ripoll, C. Gonzalez-Ballestero, G. Kirchmair, H. Ritsch, and B. Vermersch.

\clearpage
\onecolumngrid
\begin{center}

\newcommand{\beginsupplement}{%
        \setcounter{table}{0}
        \renewcommand{\thetable}{S\arabic{table}}%
        \setcounter{figure}{0}
        \renewcommand{\thefigure}{S\arabic{figure}}%
     }

\textbf{\large Supplemental Material: Cooperative Effects in Closely Packed Quantum Emitters with Collective Dephasing}
\end{center}
\twocolumngrid
\newcommand{\beginsupplement}{%
        \setcounter{table}{0}
        \renewcommand{\thetable}{S\arabic{table}}%
        \setcounter{figure}{0}
        \renewcommand{\thefigure}{S\arabic{figure}}%
     }

\setcounter{figure}{0}
\setcounter{table}{0}
\setcounter{page}{1}
\makeatletter
\renewcommand{\theequation}{S\arabic{equation}}
\renewcommand{\thefigure}{S\arabic{figure}}
\renewcommand{\bibnumfmt}[1]{[S#1]}
\renewcommand{\citenumfont}[1]{S#1}
\newcommand{\hR}{\hat{R}}
\vspace{0.8 in}

\section{Random configuration choice - Details}
Following \cite{SDamanet16}, in order to generate a random configuration with a given average separation $\bar{r}$, we first pick $N$ points uniformly distributed over a spherical container of arbitrary radius $R$. Following this, we rescale all distances by the average separation and multiply all co-ordinates by the prescribed average $\bar{r}$. If any two emitters are separated by a distance less than a cutoff (chosen as $10^{-4} \lambda_0$), the configuration is dropped. We note that the inset plot in Fig. 1 (b) of the main paper showing the mean and confidence intervals of the distribution for $\bar g \equiv \sum_{n \neq 1} \abs{g_{1n}}$ and $\bG \equiv  \sum_{n \neq 1} \G_{1n}/(N-1)$ has no qualitative dependence on which emitter we choose to index by $m=1$ within the ensemble since we are considering random distributions.

\section{Analytical calculation of $\eta$ for $N=2$}
Our method closely follows the treatment in \cite{SFicek83} that allows us to solve the $N=2$ emitter case exactly without any approximations \ie $\bG \neq \G/2$ and $\bg \neq 0$. The idea in \cite{SFicek83} is to write down the equations of motion for the operator averages $\avg{\hsig_m^{\alpha}}$, $\avg{\hsig_m^{\alpha}\hsig_n^{\beta}}$. For the $N=2$ case there are $15$ unique averages, exactly equal to the number of unique matrix elements of the density matrix. Thus solving this set of equations gives an exact solution to the master equation. Moreover, the $15$ equations also split up into two groups of $9$ and $6$ equations - one for permutation symmetric averages, such as $\avg{\hsig_1^{x}}+\avg{\hsig_2^{x}},\avg{\hsig_1^{x}\hsig_2^{y}}+\avg{\hsig_2^{x}\hsig_1^{y}}$, denoted by the column vector $\XX$, and anti-symmetric ones, such as $\avg{\hsig_1^{x}}-\avg{\hsig_2^{x}}$, denoted by the column vector $\YY$. The equation of motions have the form \cite{SFicek83}:
\begin{align}
    \frac{d \XX}{dt} &= M_X \XX + \XX_0 \label{eq:evolX}\\
    \frac{d \YY}{dt} &= M_Y \YY \label{eq:evolY}.
\end{align}
Since $M_Y$ turns out to be invertible, $\YY$ vanishes in the steady state. Thus in order to solve for the steady state we just need to invert the matrix $M_X$ in \eqnref{eq:evolX}. Interestingly, we find that the determinant of $M_X$ is proportional to $\G/2-\bG$. This means we have two distinct solutions depending on whether $\bG = \G/2$ (collective) or not \cite{SFicek83,SFicekTanasReview02}. This is so since for $\bG = \G/2$, $\hat{\mathbf{S}}^2$ is a constant of motion and one of the equations in \eqnref{eq:evolX} becomes redundant. We do not give the exact analytical expressions for $\eta$ as they are cumbersome (but have plotted the same in Fig.~2 of the main text) and consider only limiting cases which exhibit CE. We presented the result for the case with $\bG \neq \G/2$ in Eqs.~(11,12) of the main paper in the limits $\g_c/\G \rightarrow \infty$. An additional comment in this case is in order. From Fig.~2 of the main paper it is apparent that at large but finite values of $\g_c/\G$, in order to have $\eta>1$ we will need $\W_0$, even larger than the value suggested on the RHS of Eq.~(11), in order to overcome the dipole interaction induced energy shifts. In additionm we also note that in the case with perfect collective interaction $\bG = \G/2$, the equivalent expressions to Eqs. (11,12) are:
\begin{align}
\lim_{\gamma_c/\Gamma \rightarrow \infty} \pare{\eta-1} &\sim \frac{\Gamma}{4\gamma_c} \pare{\frac{\Omega_0^2}{\Gamma^2} - 2 },\label{eq:etagamcneq0coll}\\
\lim_{\Omega_0/\Gamma \rightarrow \infty,\gamma_c=0} \pare{\eta-1} &\sim \frac{1}{15} - \frac{112 \left (16 \bg^2+15 \G^2 \right)}{3315 \W_0^2} \label{eq:etagamc0coll}.
\end{align}
From the above equation, it is clear that, unlike in the case with $\bG\neq\G/2$, provided $\W_0>\sqrt{2}\G$ and $\W_0>112\sqrt{16 \bg^2+15\G^2}/225$, there is CE both in the case with and without $\g_c$.

\section{Derivation of Effective Master Equation (13)}
In order to detail the procedure for adiabatic elimination \cite{SCirac92,SLesanovsky13}, it is useful to first choose a basis that commutes with the faster Liouvillian $\mL_0$. The eigenstates of $\hS_z$ defined as:
\begin{align}
\hS_z \ket{m,\alpha_m} = 2 m \ket{m,\alpha_m} \label{eq:basis},
\end{align}
with $-N/2\leq m \leq N/2$ and $1\leq \alpha_m \leq \binom{N}{N/2+m}$ provide the required basis. Note that these are not the Dicke states. The idea then is to derive an equation for $\hmu = \mathcal{P} \rho$, where  $\mathcal{P}$ is the the projector to the dark subspace of $\mL_0$ and can be formally written as $\lim_{t \rightarrow \infty} e^{\mL_0 t}$. From the formal expression for $\mP$ we can see immediately that the projection essentially suppresses all the block off-diagonal elements of the density matrix in the basis \eqnref{eq:basis} \ie $\sandwich{m}{e^{\mL_0 t}}{n} = e^{i\D_0(m-n)t-\g_c(m-n)^2 t}$. To determine the evolution of $\hmu$, we rewrite the density matrix as $\hrho = \mu + \mQ \hrho$, with $\mQ = \mathcal{I} - \mP$ and use the full master equation to write separate equations for $\hmu$ and $\mQ \hrho$. At this point it is also convenient to split the slow Liouvillian as $\mL_1 = \mL_{1c} + \mL_{1d}$. Here $\mL_{1c} \rho= \mL_\Gamma \rho + (\im \hbar)^{-1}[\Hop_I , \hat \rho] $ commutes with $\mL_0$, and we have $\mL_{1d} \rho =  (\im \hbar)^{-1}[\hbar \Omega_0 \hat S^x, \hat \rho]$. Using the properties $\mP \mL_0 \mP = \mP \mL_0 \mQ = \mP \mL_{1d} \mP \equiv 0$, and $\mP \mL_1 \mu = \mL_{1c} \mu$, we can derive
\begin{align}
\partial_t \hmu = \mP \mL_1 \mQ \rho + \mL_{1c} \hmu. \label{eq:mueqn}
\end{align}
The key step in adiabatic elimination is to constrain the dynamics of $\mQ \hrho$ to $\hmu$ via $\mQ \hrho \approx -(\mQ \mL_1+\mQ \mL_0)^{-1} \mQ \mL_1 \hmu$ \cite{SCirac92,SLesanovsky13}, and using properties of Laplace transform to write:
\begin{align}
Q \hrho \approx \int_0^{\infty} \mQ e^{\mL_0 t} \mL_{1d} \mu,\label{eq:adbelim}
\end{align}
to first order in $\mL_{1d}$. Substituting \eqnref{eq:adbelim} in \eqnref{eq:mueqn} and using $\mP \mL_1 \mQ = \mP \mL_{1d}$, we obtain:
\begin{align}
\partial_t \hmu = \mL_{1c} \hmu + \int_0^{\infty} \mP \mL_{1d} e^{\mL_0 t} \mL_{1d} \hmu \label{eq:mueqn2}.
\end{align}
In order to simplify the above equation we note that
\begin{align}
\mP \mL_{1d} e^{\mL_0 t} \mL_{1d} \hmu &= -\frac{\W_0^2}{4} \mP \left[\hS^x e^{\mL_0 t} \hS^x \hmu -\hS^x e^{\mL_0 t} \hmu \hS^x \right . \nonumber\\
&\left . - e^{\mL_0 t} \hS^x \hmu \hS^x + e^{\mL_0 t} \hmu \hS^x \hS^x \right].\label{eq:steps1}
\end{align}
Using $\hS^x = \hS^+ + \hS^-$, and the property $\sandwich{m,\alpha_m}{\hS^{\pm}}{n,\alpha_n} \propto \delta_{m,n\pm 1}$ we can show, for example, that:
\begin{align}
\mP \hS^x e^{\mL_0 t} \hS^x \hmu = e^{-\g_c t - i\D_0 t} \hS^{+} \hS^{-} \hmu + e^{-\g_c t + i \D_0 t} \hS^{-} \hS^+ \hmu \label{eq:steps2}.
\end{align}
Constructing similar simplifications for the rest of the terms in \eqnref{eq:steps1}, we arrive at Eq. 13 of the main paper.

\section{Approximate methods}

\begin{figure}
\centering
\includegraphics[width=8cm]{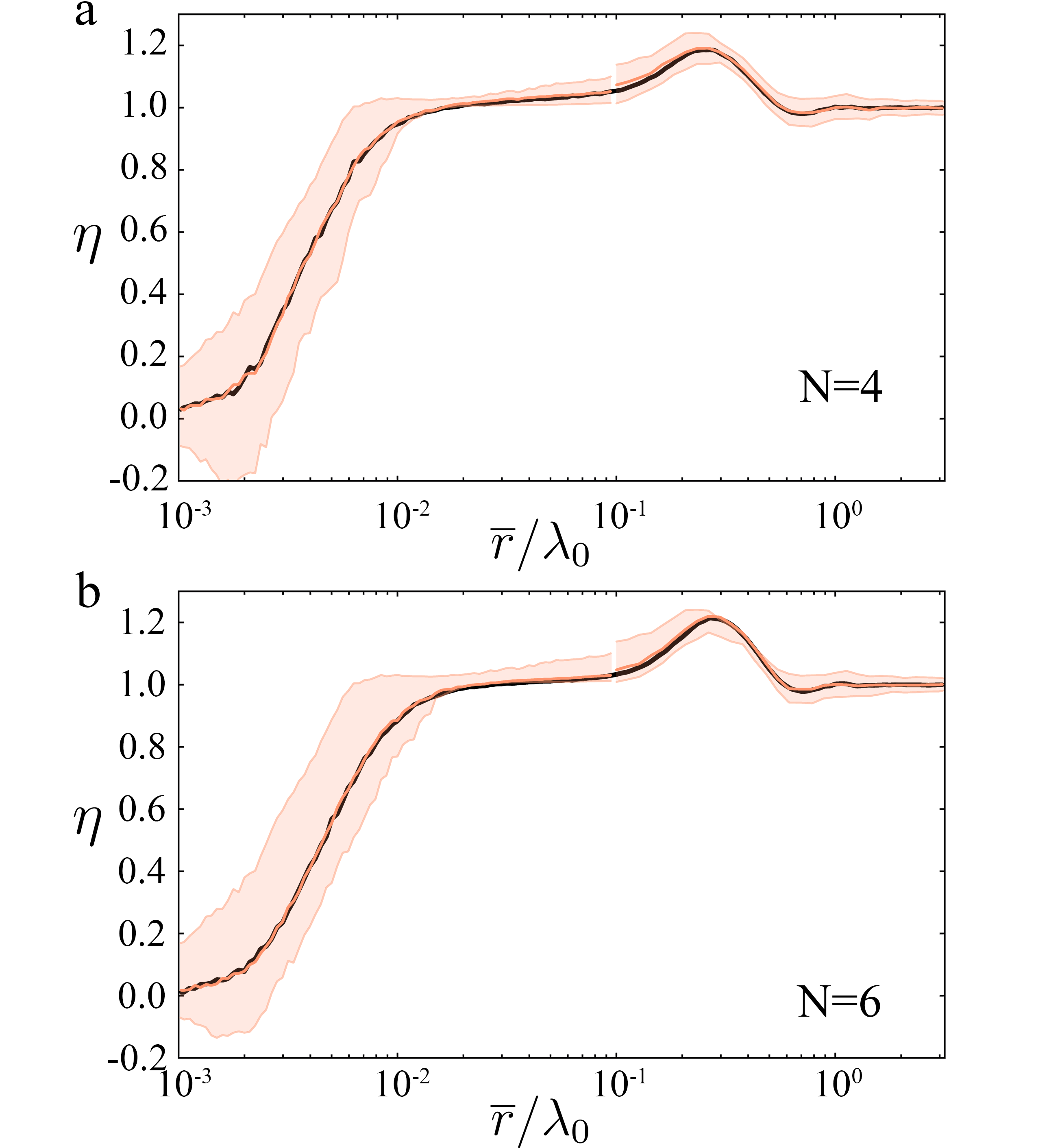}
\caption{$\avg{\eta}$ (orange lines) and statistical error (shaded region) calculated using approximation methods (see text) for random arrangements of $N=4$(a), $N=6$ (b) emitters compared to $\avg{\eta}$ (black lines) computed from the steady state of the full master equation. The other parameters are the same as in Fig. 1b.}
\label{fig:figSM_N4-6}
\end{figure}
\begin{figure}
\centering
\includegraphics[width=8cm]{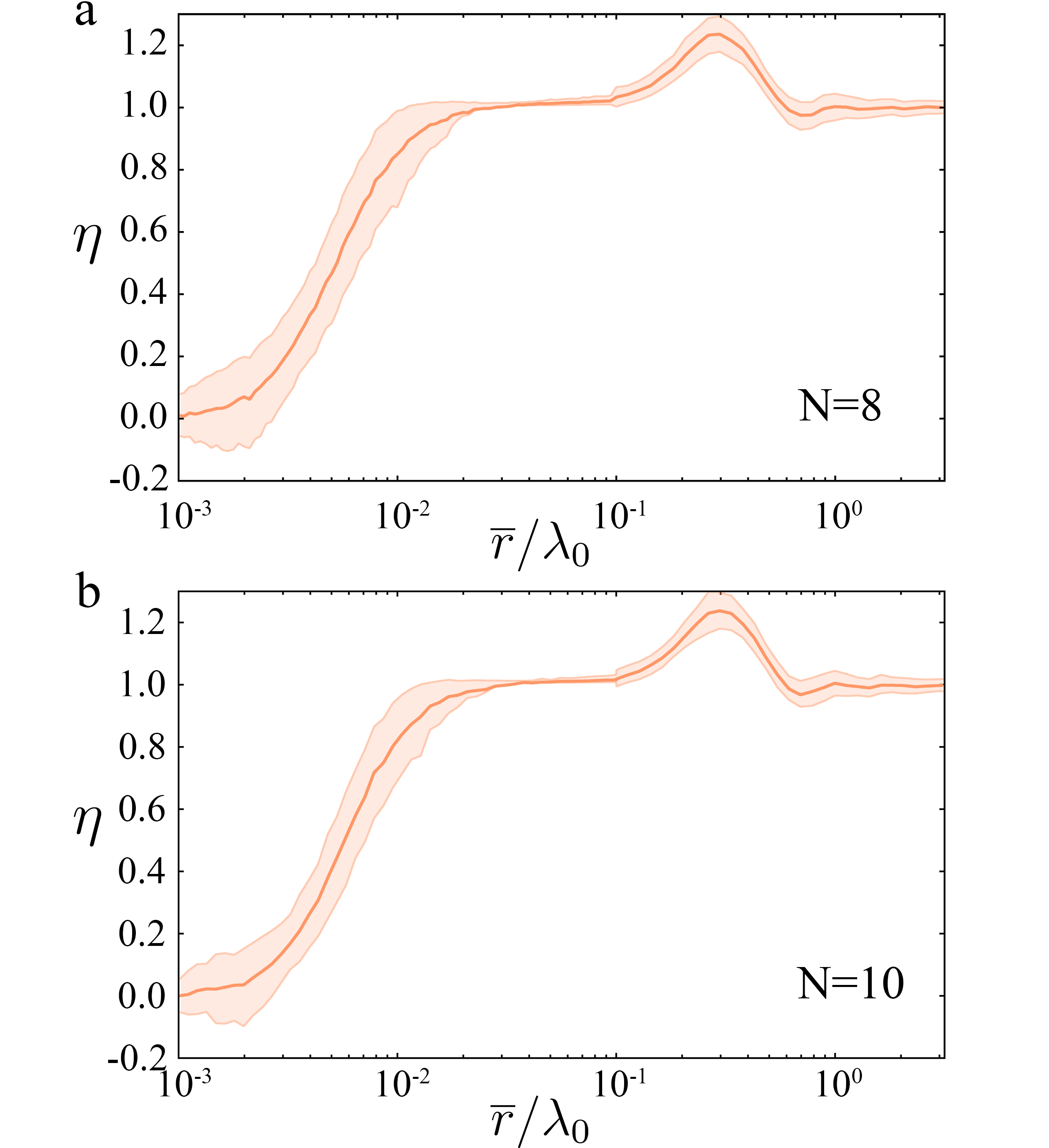}
\caption{$\avg{\eta}$ (orange lines) and statistical error (shaded region) calculated using approximation methods (see text) for random arrangements of $N=8$ (a), $N=10$ (b) emitters. The other parameters are the same as in Fig. 1b. }
\label{fig:figSM_N8-10}
\end{figure}

\figref{fig:figSM_N4-6} compares $\avg{\eta}$ obtained from approximate methods (rate equation averaged over $1000$ random configurations and trajectory simulations averaged over $200$ configurations with 500 trajectories each) for $N=4,6$ emitters to that from a direct steady state calculation for the full master equation (averaged over $1000$ configurations). The agreement is good and suggests that the approximate methods used in Fig. 3 of the main paper for larger $N$ can be trusted. The shaded region in \figref{fig:figSM_N8-10} displays the interval where $60 \%$ of $\eta$ values lie for the $N=8,10$ case. In Fig.~3 of the paper we plotted only the averaged value  $\avg{\eta}$. Note that even when $\bar{r}/\lambda_0>10^{-1}$, there are some random configurations that lead to very large $\g_c>\bg \gg \G$ leading to very slow evolution of the trajectory simulations. For such isolated configurations we calculate the steady state via a rate equation approximation to the effective master equation.

\section{Additional Numerical Results}

\begin{figure}
\centering
\includegraphics[width=8.6cm,height=5.4cm]{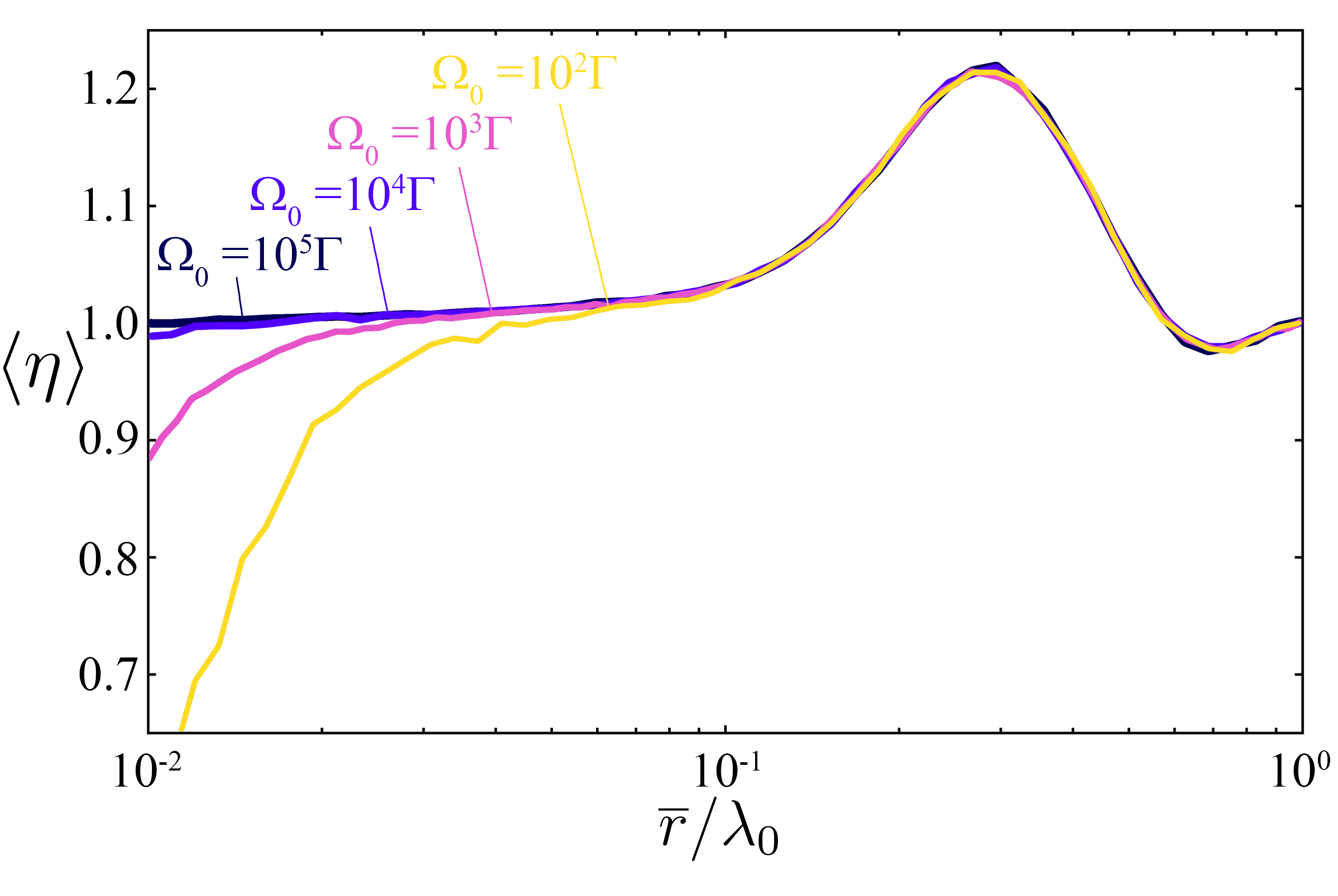}
\caption{$\avg{\eta}$ in the CE region as a function of $\W_0$ for $N=6$ randomly distributed emitters. The $\W_0 = 10^2 \G$, $\W_0 = 10^4 \G$, and $\W_0 = 10^5 \G$ cases are from an average over 500 realisations with the steady state computed for the full master equation. The other parameters are the same as in Fig. 1b.}
\label{fig:figSM_drive}
\end{figure}

\figref{fig:figSM_drive} depicts the $\avg{\eta}$ for $N=6$ randomly distributed emitters as a function of the drive strength $\W_0$. This demonstrates that in the random distribution case, increasing $\W_0$ does not lead to enhancement at smaller separations due to the strong local dephasing induced by dipole-dipole interactions. We note that the position of the CE peak is not affected by the drive strength. This is consistent with the condition obtained in the main paper by analyzing the optimal separation for two emitters within the ensemble giving $r_{12}\approx 0.2\lambda_0$, independent of the drive.

\begin{figure}
\centering
\includegraphics[width=8.6cm,height=5.4cm]{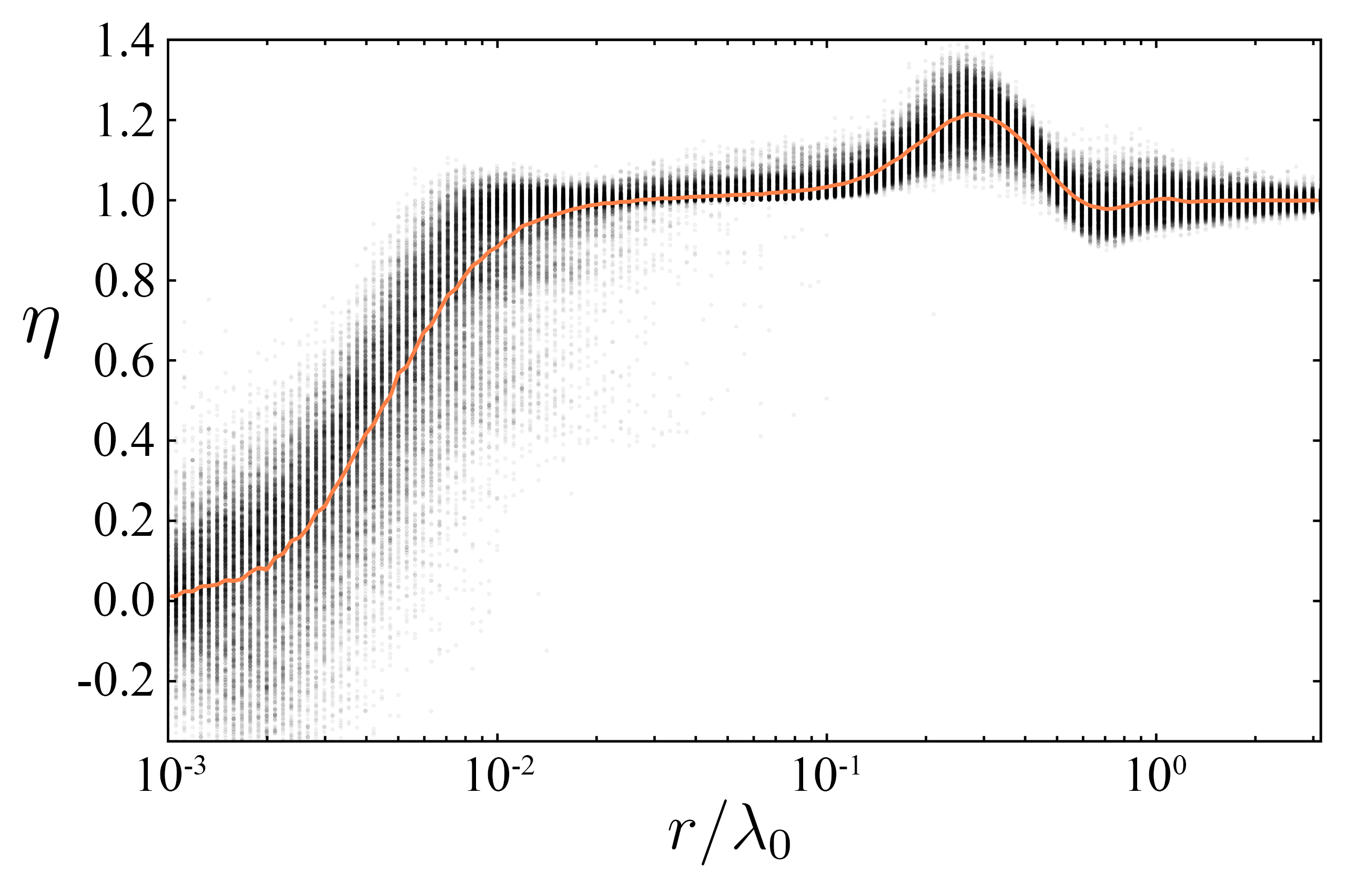}
\caption{$\eta$ for $N=6$ computed from the steady state of the full master equation for the same 1000 random configurations presented in Fig. 1b. The mean value of $\eta$ is shown by the solid line while each point represents a value of $\eta$ for a particular configuration. The parameters are the same as in Fig. 1b.}
\label{fig:figSM_N6_all}
\end{figure}

\figref{fig:figSM_N6_all} presents the numerical results for the 1000 random configurations for $N=6$. In addition to the mean value for $\eta$, already presented in Fig. 1b, we show here the values for $\eta$ for each configuration. In the region $\bar{r}/\lambda_0\in \left[10^{-2},10^{-1}\right]$, $\avg{\eta}$ is not appreciably above $1$ although a vast majority of random configurations exhibit $\eta>1$. Some configuration even show large values of $\eta$, which are likely to be related to geometrical configurations with strong symmetries (see the discussion concerning the equidistant circular arrangement). The most remarkable aspect of this figure concerns the region of maximum CE. As already mentioned, this region is consistent with the optimal density parameter anticipated from considering just $2$ emitters, $r_{12} \approx 0.2\lambda_0$, and moreover we see that \emph{all} the configurations provide CE. This suggests $\bar{r} \approx 0.2\lambda_0$ to be a fairly general criterion to maximize the CE.

\begin{figure}
\centering
\includegraphics[width=8.6cm,height=5.4cm]{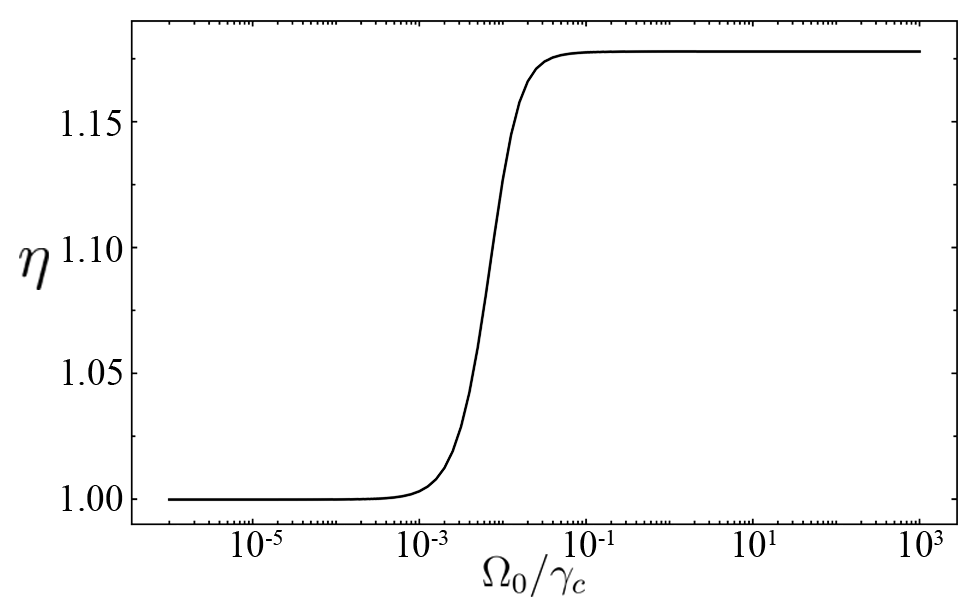}
\caption{$\eta$ for $N=6$ for a single random realization with average separation $\bar{r} = 0.2 \lambda_0$ as a function of the driving field strength $\W_0$. Other parameters are the same as in Fig. 1b.}
\label{fig:figSM_etafnw0N6}
\end{figure}

\figref{fig:figSM_etafnw0N6} illustrates the possibility of controlling $\eta$ for a given random distribution of emitters by tuning the drive strength $\W_0$. As in the case of $N=2$ emitters, depicted in Fig. 2 of the main paper, at a given value of the collective dephasing there is enhancement only when the drive strength is larger than some critical value that will in general vary from realisation to realisation.

\section{Equidistant circular arrangement of TLEs}

\begin{figure}
\centering
\includegraphics[width=8.6cm,height=5.4cm]{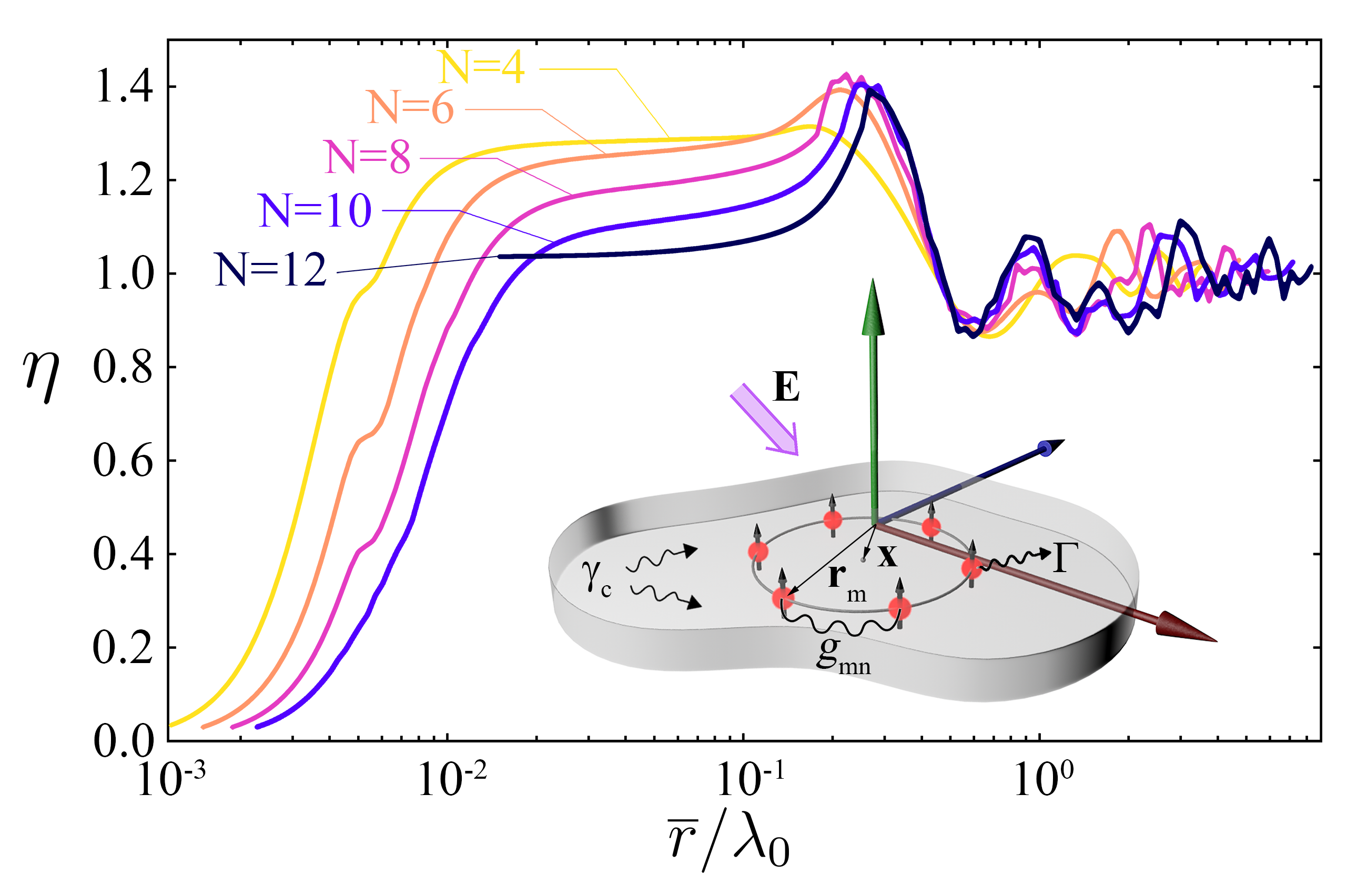}
\caption{$\eta$ for $4\leq N  \leq 12$ emitters in a circular arrangement shown schematically in the inset. $N=4,6$ results are calculated with the steady state of the full master equation. $N=8,10,12$ results are obtained using approximation methods (see text for details). The other parameters are the same as in Fig. 1b of the main paper.}
\label{fig:figSM_circle}
\end{figure}

In \figref{fig:figSM_circle}, we consider $\eta$ for an equispaced arrangement of emitters on a two dimensional circle (see inset in \figref{fig:figSM_circle} for a schematic of this configuration) as a function of the average separation $\bar{r}$. For $N=4,6$,  $\eta$ was calculated from the steady state of the full master equation without any approximation. For $N=8,10$ ($N=12$), similar to the random configuration case, we used the rate equation approximation to the full master equation (effective master equation) for small separations and trajectory simulations for larger separations. Since the effective master equation is valid only when $\bg \ll \g_c$, we do not consider very small separations for $N=12$. We can see from the results that for such a permutation symmetric configuration \cite{SGross82}, we find stronger CE that also extends over a larger region in $\bar r$ compared to the random distribution results, e.g. Fig. 1b of the main paper.

\section{Collective dephasing master equation}

We consider two situations in which a collective dephasing Markovian master equation of the form in Eq. (4) of the main paper may be justified. In the first example the TLEs of interest are NV centers in diamonds and the dephasing arises from their interaction with phonons in the diamond lattice. The interaction with phonons arises due to the modification of the energy level structure of the TLE under mechanical strain \cite{SAlbrecht13}, which in turn can be expressed in terms of lattice phonons. Let us now specialise to large wavelength acoustic phonons, that have the dispersion relation $\omega(\kk) = \omega_k = c_s \vert \kk \vert$, with $c_s = 1.2 \times 10^4$ m s$^{-1}$, the speed of sound in diamond. For the sake of simplicity, let us also restrict to just the longitudinal polarization modes parallel to the propogation momentum $\kk$. Then we obtain the following TLE-phonon interaction hamiltonian:
\begin{align}
\hat{V}_{\mathrm{ep}} &= \sum_{l} \sz_l \sum_{\kk} \sqbra{u_\kk \aop_{\kk}e^{i\kk \cdot \rr_l}+u_\kk^{*} \adop_{\kk}e^{-i\kk \cdot \rr_l}} \label{eq:SPVep},
\end{align}
with $\adop_{\kk}$ denoting the creation for a phonon with momentum $\kk$, $M$ is the mass of the crystal, $u_\kk = i  \frac{\zeta}{2} \sqrt{\frac{\hbar}{2M\omega_k}} (k_x^2+k_y^2+8\Xi^2k_z^2)/k$, and $k = \vert \kk \vert$. In the above, $\zeta \simeq 610$THz is an energy scale emerging from the strain coupling of the NV center and $\Xi$ is a factor introduced to differentiate between the Nitrogen and Carbon atoms that contribute to the coupling \cite{SAlbrecht13}. Since we are interested in a simple description capturing only the key physics, we will set $8 \Xi^2\equiv 1$ to obtain an isotropic coupling $u_\kk =u_k =  i  \frac{\zeta}{2} \sqrt{\frac{\hbar}{2M\omega_k}} k$. The above hamiltonian has the same form as the ones studied in the context of pure dephasing of qubits in quantum information \cite{SPalma96,SLidar01} and is also known as the spin-boson model. Let us now detail some of the steps towards eliminating the phononic bath and constructing a master equation \cite{SBreuerPet02,SCarmichael99} for the TLEs. First we note that for simplicity we will neglect the driving of the TLEs (we comment on this aspect in the end) and consider the following bare hamiltonian for the system:
\begin{align}
\hat{H}_0/\hbar = \sum_{\kk} \omega_k \adop_{\kk} \aop_{\kk} + \frac{\omega_0}{2} S^z \label{eq:H0ep}.
\end{align}
The interaction picture hamiltonian is then given by:
\begin{align}
\hat{V}_{\mathrm{ep}}(t) &= \sum_l \hat{R}_l(t) \sz_l \label{eq:IPVep}\\
\hat{R}_l(t) &=  \sum_{\kk} \sqbra{u_k \aop_{\kk}e^{i\kk \cdot \rr_l-i\omega_k t}+u_k^{*} \adop_{\kk}e^{-i\kk \cdot \rr_l+i\omega_k t}}\label{eq:Resops}.
\end{align}
We follow the usual Born-Markov approach to derive a master equation for the reduced density matrix of the TLEs denoted by $\hat{\rho}(t)$. To this end we assume that the total state of the system and the phonon reservoir is a product state of the form $\hat{\rho}_{tot}(t) \approx \hat{\rho}(t) \otimes \hat{\rho}_R(0)$. The reservoir is at its initial state, chosen as a thermal state $\hat{\rho}_R(0) = \exp\pbra{-\beta \hbar \sum_{\kk} \omega_k \adop_{\kk} \aop_{\kk} }$ with $\beta = 1/(k_B T)$, and is essentially affected very little by the system at all times. This is the content of the Born approximation, which is valid here since the typical magnitude of $u_k \sim 2 \pi \times 0.3 $ MHz (for a cubic diamond sample of size $\lambda_0$) and is much less than the optical transition frequency of the NV center. In order to justify the validity of the Markov approximation, we have to examine the two-time correlations of the reservoir operators $\hR_l(t)$. For this, it is essential to choose a more specific model for the phonon density of states. For the sake of simplicity, we consider a cubical region of finite volume $V = L^3$ with periodic boundary conditions along all three directions. This immediately leads to the following allowed values of the wavenumbers \cite{SKittel96}:
\begin{align}
k_x,k_y,k_z = 0,\pm \frac{2 \pi}{L},\pm \frac{4 \pi}{L}, \dots \,\, .
\end{align}
We will assume the Debye model and introduce a cut-off frequency $\omega_D = c_s (6 \pi^2 n)^{1/3}$ with $n$ the number density of diamond \cite{SKittel96}. The density of states is then given by: 
\begin{align}
D(\omega_k) = \begin{cases} 
\frac{V \omega_k^2}{c_s^3 (2 \pi)^2},& \mathrm{when}\, \omega_k \leq \omega_D.\\
0, & \mathrm{when}\, \omega_k > \omega_D.
\end{cases}
\end{align}
It is then straight forward to evaluate the correlation function $R_{ml} = \avg{\hR_m(0) \hR_l(\tau)}$ and it can be shown that it decays over the time scale $1/\omega_T$, where $\omega_T = k_B T/\hbar$ \cite{SCarmichael99}. Thus the Markovian approximation is justified as long as the system dynamics is over time scales slower than $1/\omega_T$. The master equation within the Born-Markov approximation then reads:
\begin{align}
\frac{d \rho}{dt} = \sum_{lm} &\frac{\g_{ml}}{2} \pbra{2\sz_l\rho \sz_m-\rho \sz_m \sz_l - \sz_m \sz_l \rho} \nonumber \\
& - i \sqbra{-S_{ml} \sz_m \sz_l,\rho} \label{eq:markovdepheqn},
\end{align}
with $\g_{ml} = R_{ml}+R_{lm}^{*}$ representing a positive definite dephasing matrix $\gamma$ and $S_{ml}=\pbra{R_{ml}-R_{lm}^{*}}/(2i)$ representing the hermitian Lamb-shift matrix \cite{SBreuerPet02}. The explicit for the dephasing rate is given by:
\begin{align}
\g_{ml} \approx \lim_{\omega_k \rightarrow 0} \frac{VG_0}{2\pi c_s^3 }\omega_k^3 \frac{\sin(\omega_k \tilde{r}_{ml})}{\omega_k \tilde{r}_{ml}} \frac{k_B T}{\hbar \omega_k} \label{eq:gmlzerolimit},
\end{align}
with $G_0 = \hbar \zeta^2/(8 M c_s^2)$, which gives a measure of the NV center phonon coupling, and $\tilde{r}_{ml} = r_{ml}/c_s$. If we naively apply the limit above it is clear that we get $\g_{ml} = 0$. But, since we are in a finite crystal with periodic boundary condition and the allowed phonon mode of lowest frequency has a wavenumber $k_f = 2 \pi/L = \omega_f/c_s$, we impose an infra-red cut off to the limit and evaluate the limit \eqnref{eq:gmlzerolimit} at $\omega_k = \omega_f$ to get:
\begin{align}
\g_{ml} \approx 4 \pi^2 G_0 \frac{k_B T}{\hbar \omega_f} \frac{\sin(k_f r_{ml})}{k_f r_{ml}}  \label{eq:gmlinfraredcutoff}.
\end{align}
If the average emitter separation is much smaller compared to the fundamental mode of the diamond, the factor $\sin(k_f r_{ml})/k_f r_{ml} \sim 1$ (for example this factor is about $0.93$ for $k_f r_{ml} \approx 0.1 \times 2 \pi$) and $\g_{ml} = \g_c$ and the dephasing becomes collective, with:
\begin{align}
\gamma_c \approx 4 \pi^2 G_0 \frac{k_B T}{\hbar \omega_f}, \label{eq:gamcfinal}.
\end{align}
This is intuitive as it is simply given by the frequency $G_0$ characterizing the NV center phonon coupling times the occupation number $k_B T/ (\hbar \omega_f)$ of the lowest phonon mode. It is also clear that the Markov approximation is valid since $\gamma_c/\omega_T = 4 \pi^2 G_0/\omega_f$ and is of the order of $10^{-8}$ for a diamond sample with $L  = \lambda_0$ (the optical wavelength). For this case, we also get a collective dephasing rate of $\gamma_c = 2 \pi \times 0.1$ MHz at room temperature. We note that this collective dephasing rate is small compared to the spontaneous emission rate of the NV center optical transition and not as large as the values we use in the main paper. This is a limitation of the rather simple two-level model of the quantum emitter we have used to describe phonon induced dephasing. In priniciple, the excited state of NV centers are multiplets and phonon interactions could mediate coupling between such excited state multiplets. This can then lead to large dephasing seen in experiments which also scales very differently with temperature (see \cite{SFu2009}). Finally, we should add that the vanishing of the dephasing rate when we do not impose the infra-red cut-off is very specific to the dimensionality of space and the behaviour of the coupling $u_k \propto \sqrt{k}$. For instance in 1-D, we can easily check that the limit $\omega_k \rightarrow 0$ leads to the finite dephasing rate:
\begin{align}
\g_{ml}^{1d} = \frac{G_0}{2}  \frac{k_B T}{\hbar \omega_f} \lim_{\omega_k \rightarrow 0}\cos(\omega_k \tilde{r}_{lm}).\label{eq:gmlin1d}
\end{align} 
For the sake of completeness, we note that the Lamb-shift term in \eqnref{eq:markovdepheqn} is given by:
\begin{align}
S_{ml} = \frac{G_0}{2 \pi^2}\frac{V}{r_{ml}^3} \sqbra{\sin \pbra{k_D r_{ml}}-k_Dr_{ml} \cos \pbra{k_D r_{ml}}} \label{eq:lambshiftdephint},
\end{align}
with the Debye wavenumber $k_D = \omega_D/c_s$. For average emitter separations of $r_{ml} \sim \lambda_0/10$ in a sample of size $L \sim \lambda_0$, the magnitude of $S_{ml} \approx 2 \pi \times 0.4$ MHz.  Since this term also breaks the collective symmetry of the system, let us compare this to the dipole-dipole interaction strength $\bg$.  For $r_{ml} \sim \lambda_0/10$, $\bg \sim \G$, spontaneous emission rate of the NV optical transition which is at least one order of magnitude larger than $S_{ml}$. Nonetheless, it would be interesting to study systematically how this affects collective enhancement as part of future work. 

Finally, we would like to point out another mechanism to generate collective dephasing. Consider a collection of TLEs that are susceptible to externally applied magnetic fields with a general hamiltonian of the form:
\begin{align}
H_{B} = B(t) \sum_l p_l \sz_l \label{eq:appliedB}.
\end{align}
As detailed in the appendix A of \cite{SDorner12}, if the applied field is of the white noise type, namely $B(t) = \sqrt{\frac{\g_c}{2}} \xi(t)$ with $\avg{\xi(t)} = 0$ and $\avg{\xi(t) \xi(t^{\prime})} = \delta(t-t^{\prime})$ and has a large correlation length in space \ie $p_l \equiv 1$, then it leads to a collective dephasing master equation of the form Eq. (4) in the main paper.

\end{document}